\documentclass[aps,twocolumn,prd,tightenlines,superscriptaddress,showpacs,byrevtex]{revtex4}
\usepackage{epsfig, amssymb}

\begin{document}
\advance\hoffset by  -4mm

\newcommand{\de}{\Delta E}
\newcommand{\mbc}{M_{\rm bc}}
\newcommand{\bb}{B{\bar B}}
\newcommand{\qq}{q{\bar q}}
\newcommand{\ks}{K_S}
\newcommand{\kpi}{K^+\pi^-}
\newcommand{\kpipin}{\kpi\pi^0}
\newcommand{\kpipipi}{\kpi\pi^+\pi^-}
\newcommand{\kpipi}{K^-\pi^+\pi^+}
\newcommand{\kspi}{\ks\pi^+}
\newcommand{\dndp}{D^+ \bar{D}^0}
\newcommand{\dndm}{D^- D^0}
\newcommand{\dndb}{D^0 \bar{D}^0}
\newcommand{\bdndp}{B^+\to\dndp}
\newcommand{\bdndm}{B^-\to\dndm}
\newcommand{\bdndb}{B^0\to\dndb}
\newcommand{\brdndp}{(3.85\pm 0.31\pm 0.38)\times 10^{-4}}
\newcommand{\brdndb}{0.43\times 10^{-4}}
\newcommand{\acpnbdp}{0.00\pm 0.08 \pm 0.02}
\newcommand{\acpul}{-0.14<A_{CP}<0.14}
\newcommand{\br}{{\mathcal B}}

\title{\Large \rm\quad\\[0.5cm] 
Measurement of the branching fraction and charge asymmetry of the decay 
$\bdndp$ and search for $\bdndb$
}

\affiliation{Budker Institute of Nuclear Physics, Novosibirsk}
\affiliation{Chiba University, Chiba}
\affiliation{University of Cincinnati, Cincinnati, Ohio 45221}
\affiliation{Gyeongsang National University, Chinju}
\affiliation{Hanyang University, Seoul}
\affiliation{University of Hawaii, Honolulu, Hawaii 96822}
\affiliation{High Energy Accelerator Research Organization (KEK), Tsukuba}
\affiliation{Institute of High Energy Physics, Chinese Academy of Sciences, Beijing}
\affiliation{Institute of High Energy Physics, Vienna}
\affiliation{Institute of High Energy Physics, Protvino}
\affiliation{Institute for Theoretical and Experimental Physics, Moscow}
\affiliation{J. Stefan Institute, Ljubljana}
\affiliation{Kanagawa University, Yokohama}
\affiliation{Korea University, Seoul}
\affiliation{Kyungpook National University, Taegu}
\affiliation{\'Ecole Polytechnique F\'ed\'erale de Lausanne (EPFL), Lausanne}
\affiliation{University of Maribor, Maribor}
\affiliation{Nagoya University, Nagoya}
\affiliation{Nara Women's University, Nara}
\affiliation{National Central University, Chung-li}
\affiliation{National United University, Miao Li}
\affiliation{Department of Physics, National Taiwan University, Taipei}
\affiliation{H. Niewodniczanski Institute of Nuclear Physics, Krakow}
\affiliation{Nippon Dental University, Niigata}
\affiliation{Niigata University, Niigata}
\affiliation{Osaka City University, Osaka}
\affiliation{Osaka University, Osaka}
\affiliation{Panjab University, Chandigarh}
\affiliation{Saga University, Saga}
\affiliation{University of Science and Technology of China, Hefei}
\affiliation{Seoul National University, Seoul}
\affiliation{Sungkyunkwan University, Suwon}
\affiliation{University of Sydney, Sydney, New South Wales}
\affiliation{Toho University, Funabashi}
\affiliation{Tohoku Gakuin University, Tagajo}
\affiliation{Department of Physics, University of Tokyo, Tokyo}
\affiliation{Tokyo Institute of Technology, Tokyo}
\affiliation{Tokyo Metropolitan University, Tokyo}
\affiliation{Tokyo University of Agriculture and Technology, Tokyo}
\affiliation{Virginia Polytechnic Institute and State University, Blacksburg, Virginia 24061}
\affiliation{Yonsei University, Seoul}
   \author{I.~Adachi}\affiliation{High Energy Accelerator Research Organization (KEK), Tsukuba} 
   \author{H.~Aihara}\affiliation{Department of Physics, University of Tokyo, Tokyo} 
   \author{K.~Arinstein}\affiliation{Budker Institute of Nuclear Physics, Novosibirsk} 
   \author{V.~Aulchenko}\affiliation{Budker Institute of Nuclear Physics, Novosibirsk} 
   \author{T.~Aushev}\affiliation{\'Ecole Polytechnique F\'ed\'erale de Lausanne (EPFL), Lausanne}\affiliation{Institute for Theoretical and Experimental Physics, Moscow} 
   \author{A.~M.~Bakich}\affiliation{University of Sydney, Sydney, New South Wales} 
   \author{V.~Balagura}\affiliation{Institute for Theoretical and Experimental Physics, Moscow} 
   \author{K.~Belous}\affiliation{Institute of High Energy Physics, Protvino} 
   \author{V.~Bhardwaj}\affiliation{Panjab University, Chandigarh} 
   \author{U.~Bitenc}\affiliation{J. Stefan Institute, Ljubljana} 
% ...
\collaboration{The Belle Collaboration}

\noaffiliation

\begin{abstract}
We report an improved measurement of the
$B^+\to D^+\bar{D}^0$ and $B^0\to D^0\bar{D}^0$
decays based on $657\times 10^6$ $\bb$ events
collected with the Belle detector at KEKB.
We measure the branching fraction and charge asymmetry for the $\bdndp$
decay: $\br(\bdndp)=\brdndp$ and $A_{CP}(\bdndp)=\acpnbdp$,
where the first error is statistical and the second is systematic.
We also set the upper limit for the $\bdndb$ decay:
$\br(\bdndb)<\brdndb$ at 90\% CL.
\end{abstract}
\pacs{13.25.Hw, 14.40.Lb}
\maketitle

Recently, evidence of direct $CP$ violation in $B^0\to D^+ D^-$ decays was
observed by the Belle collaboration~\cite{belle_dpdm}, while BaBar measured 
an asymmetry consistent with zero~\cite{babar_dpdm}. A large direct $CP$ 
asymmetry for this decay can indicate the presence of new-physics effects 
in the electroweak penguin sector~\cite{fleisher}.
A similar effect might be observable in the charged mode $\bdndp$. 
This decay has already been observed by Belle~\cite{belle_dndp} 
and confirmed by BaBar~\cite{babar_dndp}. The latter analysis also includes 
charge asymmetry measurement.

In this paper, we report an improved measurement of
the branching fraction and charge asymmetry for $\bdndp$ decay and
a search for the decay $\bdndb$.
The latter can only be produced by a $W$ exchange diagram.
We use a data sample of $(657\pm 9)\times 10^6$ $\bb$ events
collected with the Belle detector at the KEKB collider~\cite{KEKB}.
The inclusion of
charge-conjugate states is implicit throughout this paper.


The Belle detector is a large-solid-angle magnetic
spectrometer that consists of a silicon vertex detector (SVD),
a 50-layer central drift chamber (CDC), an array of
aerogel threshold Cherenkov counters (ACC),
a barrel-like arrangement of time-of-flight
scintillation counters (TOF), and an electromagnetic calorimeter (ECL)
comprised of CsI(Tl) crystals located inside
a superconducting solenoid coil that provides a 1.5~T
magnetic field.  An iron flux-return located outside
the coil is instrumented to detect $K_L$ mesons and to identify
muons (KLM).  The detector is described in detail elsewhere~\cite{NIM}.
For the first sample
of 152 million $B\bar{B}$ pairs, a 2.0 cm radius beam pipe
and a 3-layer silicon vertex detector were used;
for the latter 505 million $B\bar{B}$ pairs,
a 1.5 cm radius beam pipe, a 4-layer silicon detector
and a small-cell inner drift chamber were used~\cite{Ushiroda}.

Each track's transverse momentum with respect to the beam axis is required 
to be greater than 0.075~GeV$/c$ in order to reduce the combinatorial 
background. 
For charged particle identification (PID), 
the measurement of the specific ionization ($dE/dx$) in the CDC,
and signals from the TOF and the ACC are used.
Charged kaons are selected with PID criteria that have
an efficiency of 88\% with a pion misidentification probability of 8\%.
All charged tracks that are consistent with a pion
hypothesis and that are not positively identified as electrons are 
treated as pion candidates.

Neutral kaons are reconstructed in the decay $\ks\to\pi^+\pi^-$;
no PID requirements are applied for the daughter pions.
The two-pion invariant mass is required to be within 9~MeV$/c^2$
($\sim 3\sigma$) of the $K^0$ mass and the displacement of the 
$\pi^+\pi^-$ vertex from the interaction point (IP) in the transverse 
($r-\varphi$) plane is required to be between 0.2~cm and 20~cm. 
The $\ks$ momentum and the vector from the IP to the
$\pi^+\pi^-$ vertex are required to be collinear in the $r-\varphi$ 
plane to within 0.2 radians.

Photon candidates are selected from ECL showers not associated
with charged tracks.
An energy deposition of at least 75~MeV and a photon-like shape 
of the shower are required for each candidate.
A pair of photons with an invariant mass 
within 12~MeV$/c^2$ ($\sim 2.5\sigma$) of the $\pi^0$ mass is 
considered as a $\pi^0$ candidate.
We require that the $\pi^0$ momentum be greater than 0.35~GeV$/c$ in order 
to reduce the combinatorial background.

We reconstruct $\bar{D}^0$ mesons in the $\kpi$, $\kpipipi$ and 
$\kpipin$ decay channels. The $D^+$ candidates are reconstructed in
the $K^-\pi^+\pi^+$ and $\ks\pi^+$ final states.
We require the invariant mass of the $\bar{D}^0$ and $D^+$ candidates to 
be within 11~MeV$/c^2$ ($1.5\sigma$ for $\kpipin$ and 
$2.5\sigma$ for other modes) of their nominal mass.
We perform a mass-constrained fit for $D$ candidates to improve their 
momentum resolution.

To suppress the large background from $B^+\to D_s^+\bar{D}^0$ with the $K^+$ 
from the $D_s^+$ decay misidentified as a pion, none of the
pions from $D^+$ should be consistent with the kaon hypothesis.
This requirement has an efficiency of 93\% and kaon misidentification 
probability of 9\%.

We combine $\bar{D}^0$ and $D^+$ ($D^0$) candidates to form $B^+$ ($B^0$) 
candidates.
These are identified by their center-of-mass (CM) energy difference, 
\mbox{$\de=(\sum_iE_i)-E_{\rm beam}$}, and the beam constrained mass, 
$\mbc=\sqrt{E_{\rm beam}^2-(\sum_i\vec{p}_i)^2}$, where $E_{\rm beam}$ 
is the beam energy and $\vec{p}_i$ and $E_i$ are the momenta and 
energies of the decay products of the $B$ meson in the CM frame. 
We select events with $\mbc>5.2$~GeV$/c^2$
and $|\de|<0.3$~GeV, and define a $B$ signal region of $|\de|<0.02$~GeV,
$5.273$~GeV$/c^2<\mbc<5.287$~GeV$/c^2$.
In an event with more than one $B$ candidate, we choose the one
with smallest $\chi^2$ from the $D$ mass-constrained fit.
We use Monte Carlo (MC) simulation to model the response of
the detector and determine the efficiency~\cite{GEANT}.

Variables that characterize the event topology are used to suppress 
background from the jet-like $e^+e^-\to\qq$ continuum process.
We require $|\cos\theta_{\rm thr}|<0.8$, where $\theta_{\rm thr}$ is 
the angle between the thrust axis of the $B$ candidate and that of the 
rest of the event; this condition rejects 77\% of the continuum 
background while retaining 78\% of the signal. 
To suppress high background in the $\dndb$ final state, we use a
Fisher discriminant, ${\cal F}$, that is based on the production
angle of the $B$ candidate, the angle of the $B$ candidate thrust axis
with respect to the beam axis, and nine parameters that characterize
the momentum flow in the event relative to the $B$ candidate thrust
axis in the CM frame~\cite{VCal}. We impose a requirement on
${\cal{F}}$ that rejects 52\% of the remaining continuum background
and retains 86\% of the signal.


\begin{table*}
\caption{Yields from the $\de$, $\mbc$ and 2D ($\de$-$\mbc$) fits, 
detection efficiencies including intermediate branching fractions, 
and corresponding branching fractions.
Upper limits are at the 90\% CL.}
\medskip
\label{defit}
\begin{tabular*}{\textwidth}{l@{\extracolsep{\fill}}ccccc}\hline\hline
Decay channel & $\de$ yield & $\mbc$ yield & 2D yield & $\varepsilon$, $10^{-4}$& 
${\cal B}$, $10^{-4}$\\\hline

$\bdndb$ & $-4.5\pm 29.7$ & $5.7\pm 28.6$ & $0.4\pm 24.8$ ($<46$)& 
$16.4$ & $<0.43$ \\

$B^\pm\to D^\pm D^0$ & $366.4\pm 31.8$ & $376.4\pm 30.7$ & $369.7\pm 29.4$ & 
$14.6$ & $3.85\pm 0.31\pm 0.38$ \\
\hline\hline
\end{tabular*}
\end{table*}

We determine the signal yield from the two-dimensional unbinned extended 
maximum likelihood fit (2D) to 
the $\de$-$\mbc$ distribution.  The signal probability density function 
(PDF) is described by a double Gaussian for $\de$ and a single Gaussian for 
$\mbc$, taking into account $\de$-$\mbc$ correlations.
We use the $B^+\to D_s^+\bar{D}^0$ events in our data sample to calibrate 
the means and resolutions of the signal shape.
The continuum, $\bb$ and $B^+\to D_s^+\bar{D}^0$ background contributions are
described separately.
We use a linear function for $\de$ and a threshold function for 
$\mbc$~\cite{argus} to describe the continuum PDF.
The $\bb$ background is modeled by a quadratic polynomial for $\de$, 
a threshold function for $\mbc$ combined with a small peaking component 
(a wide Gaussian for $\de$ and a Gaussian for $\mbc$). 
The shape of the peaking background and parameters of the threshold 
function are fixed from the generic $\bb$ MC.
The $\de$ linear slope and quadratic term are free parameters.
The peak in the $\de$ distribution near $-70$ MeV coming from the 
$B^+\to D_s^+ \bar{D}^0$ decay is described by
a Gaussian for $\de$ and a Gaussian for $\mbc$.
Again, we use $B^+ \to D_s^+ \bar{D}^0$ to obtain the
parameters of this PDF.
The region $\de<-0.1$~GeV is excluded from the fit 
to avoid contributions from $B\to\bar{D} D^*$ decays.

\begin{figure*}
  \includegraphics[width=0.39\textwidth] {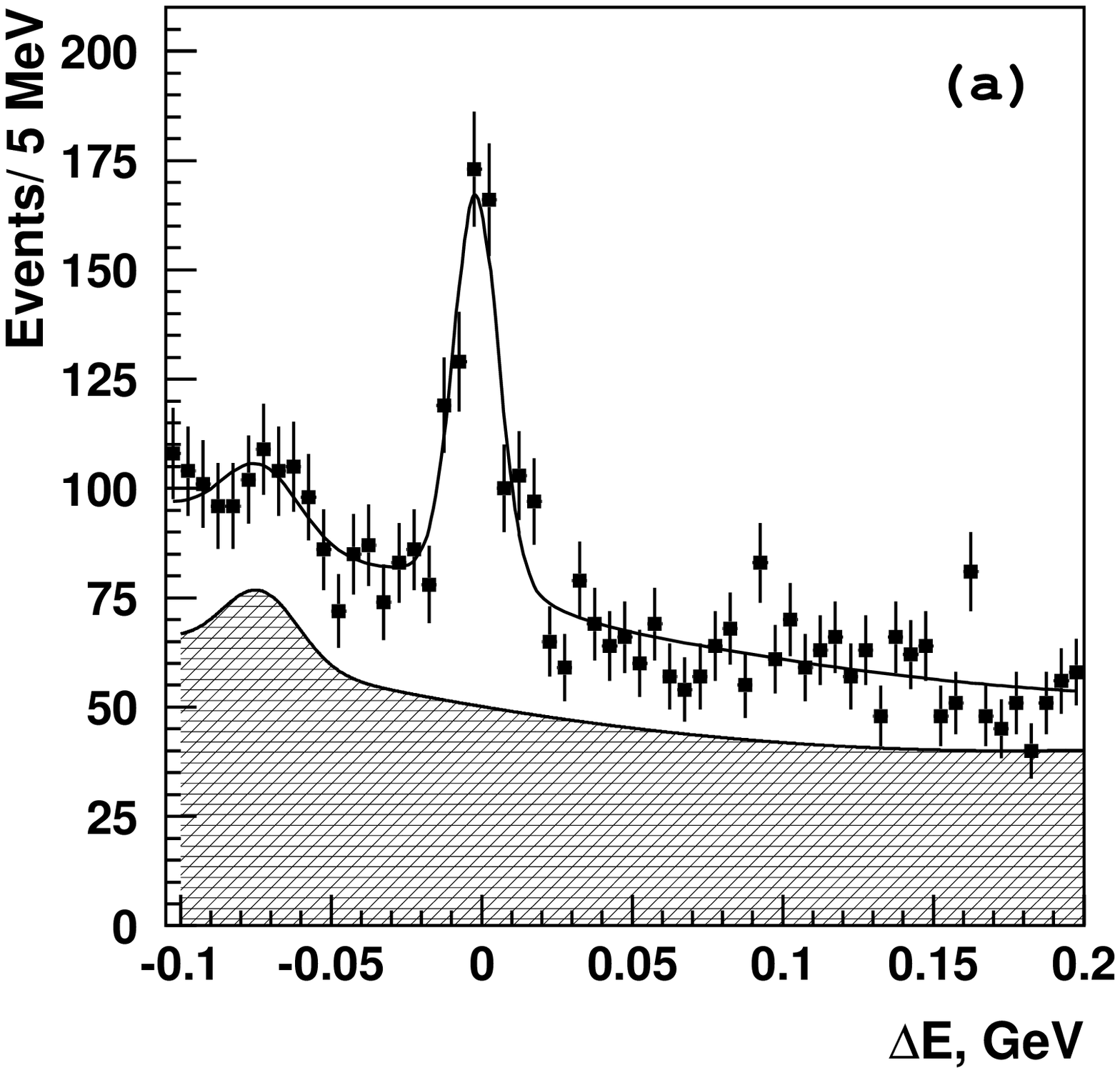} \hfill
  \includegraphics[width=0.39\textwidth] {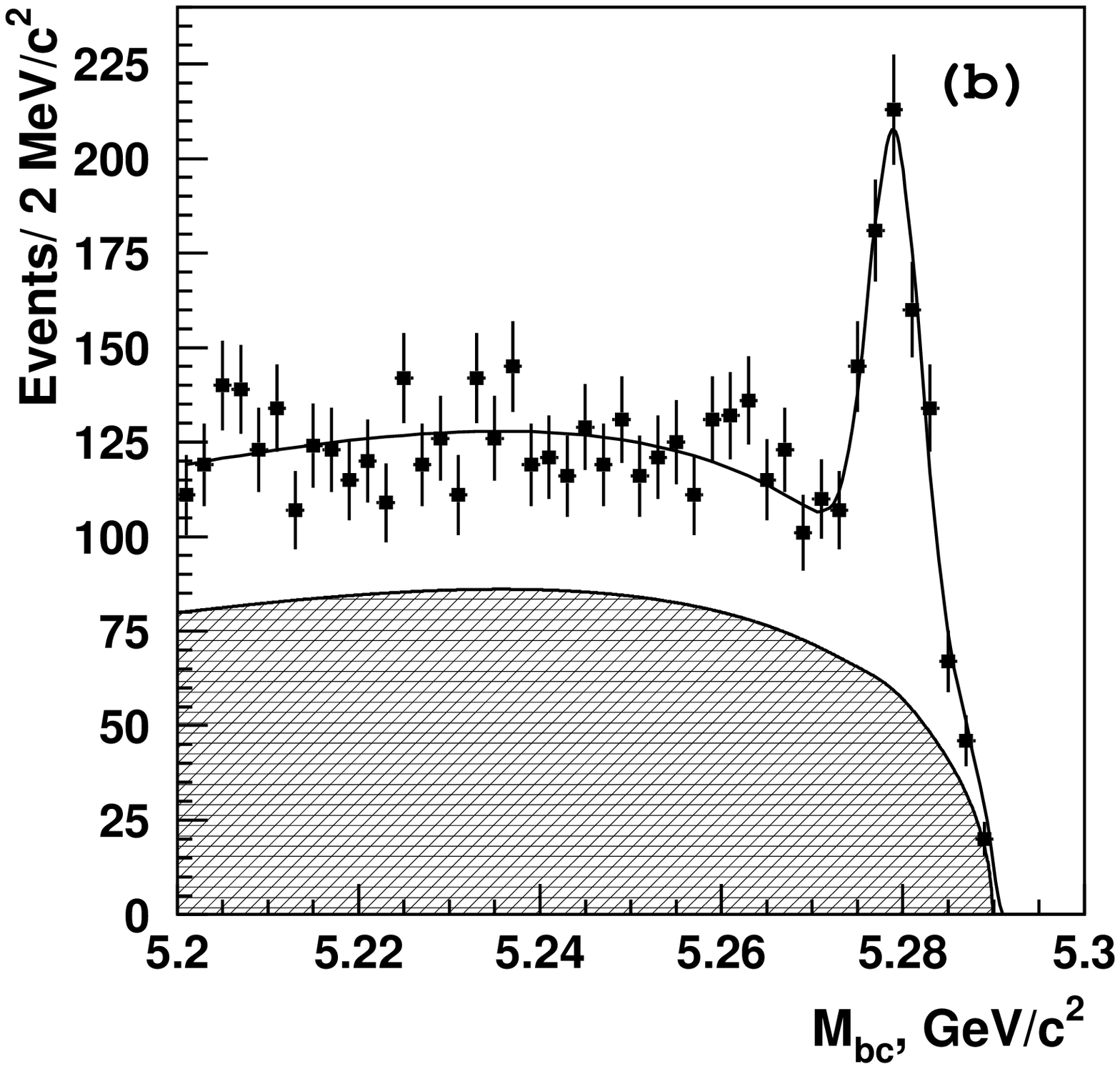}
  \caption{$\de$ (a) and $\mbc$ (b) distributions for the
    $\bdndp$ candidates. Each distribution is the projection of the 
    signal region of the other parameter. Points with errors
    represent the experimental data, open curves show projections from the 
    2D fits and cross-hatched curves show the $\bb$ component only.}
  \label{dndc_mbcde}
\end{figure*}

\begin{figure*}
  \includegraphics[width=0.39\textwidth] {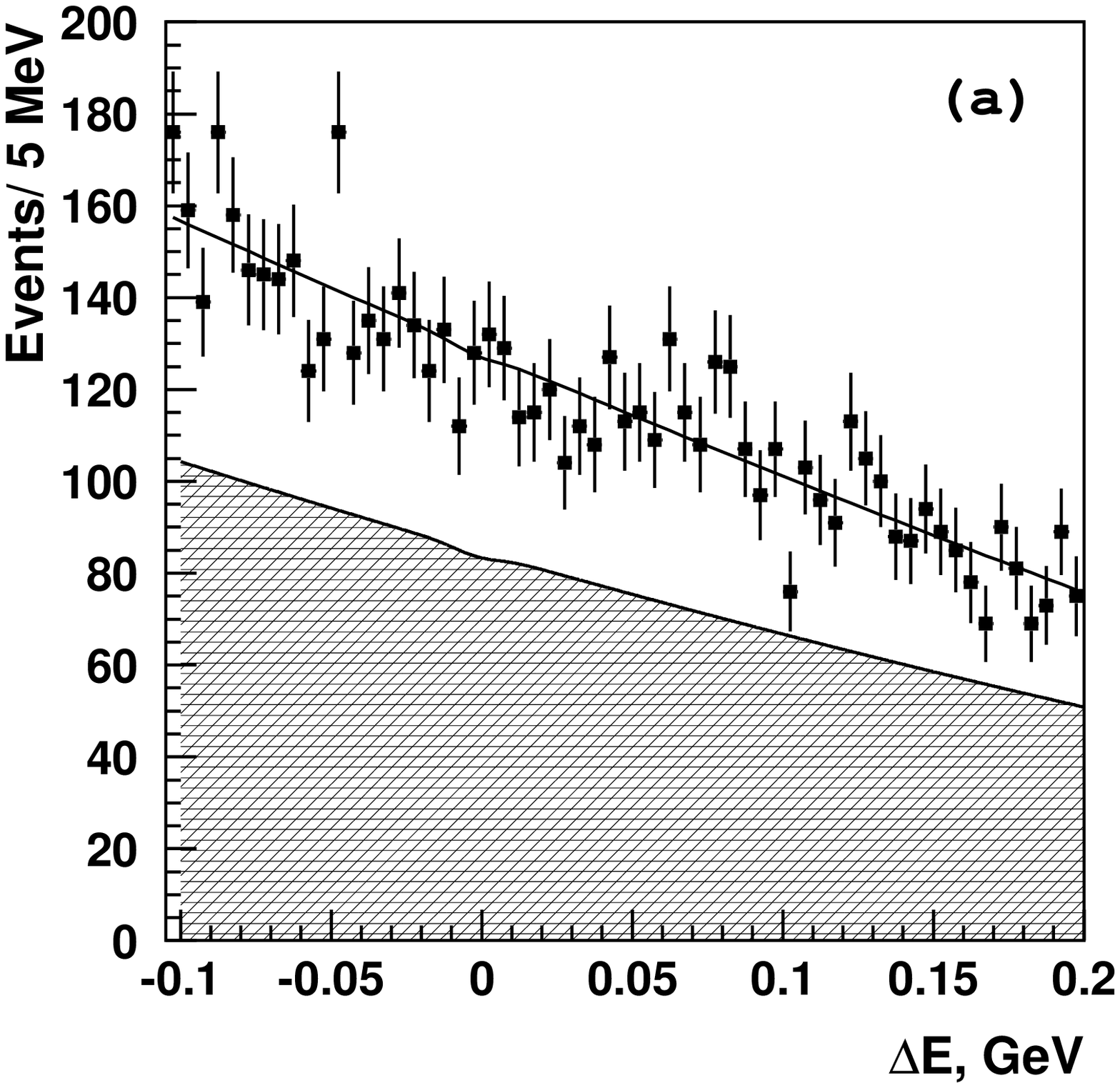} \hfill
  \includegraphics[width=0.39\textwidth] {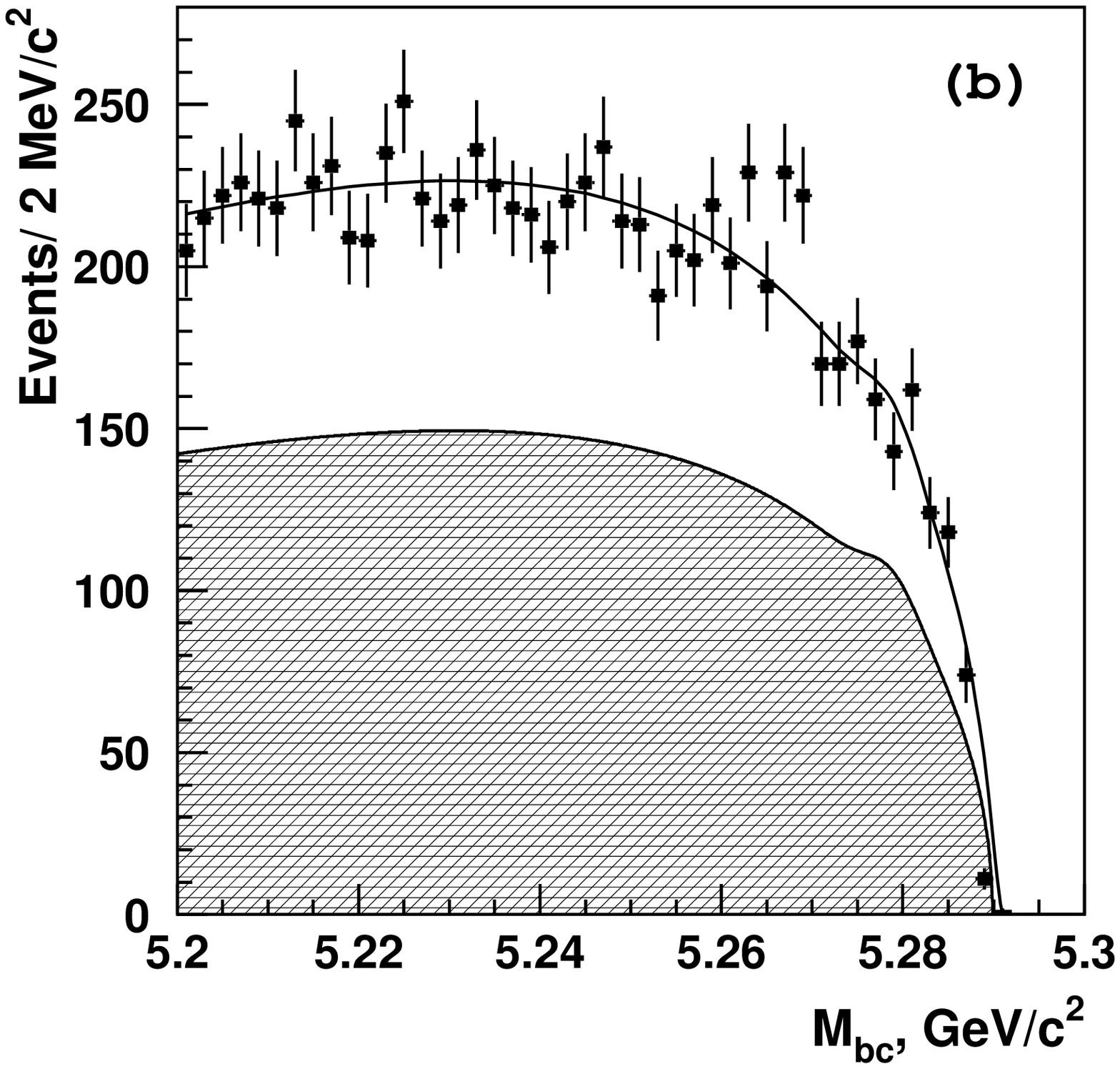}
  \caption{$\de$ (a) and $\mbc$ (b) distributions for the
    $\bdndb$ candidates. Each distribution is the projection of the 
    signal region of the other parameter. Points with errors
    represent the experimental data, open curves show projections from the 
    2D fits and cross-hatched curves show the $\bb$ component only.}
  \label{dndb_mbcde}
\end{figure*}

As a cross-check, we also perform separate one-dimensional fits to the $\de$ 
and $\mbc$ distributions, requiring the other variable to be in the 
signal region. The results are given in Table~\ref{defit}, 
where the listed efficiencies include intermediate branching fractions.
The projections of the 2D fit result are shown in
Figs.~\ref{dndc_mbcde} and \ref{dndb_mbcde}.
The 90\% confidence level (CL) upper limit for $\bdndb$ signal yield is 
obtained using the POLE program~\cite{pole} based on the 
Feldman-Cousins method~\cite{cite:FC}. The systematic uncertainty 
(described later) is taken into account in this calculation.

\begin{table*}
\caption{Charged $B$ meson yields from the $\de$, $\mbc$ and 
2D ($\de$-$\mbc$) fits. Errors are statistical only.}
\medskip
\label{asymfit}
\begin{tabular*}{\textwidth}{l@{\extracolsep{\fill}}ccc}\hline\hline
Decay channel & $\de$ yield & $\mbc$ yield & 2D yield\\\hline
$\bdndp$ & $183.9\pm 21.5$ & $184.4\pm 21.4$ & $184.2\pm 20.4$\\
$\bdndm$ & $183.4\pm 22.1$ & $192.5\pm 21.8$ & $185.4\pm 21.0$\\
\hline\hline
\end{tabular*}
\end{table*}

\begin{figure*}
  \includegraphics[width=0.39\textwidth] {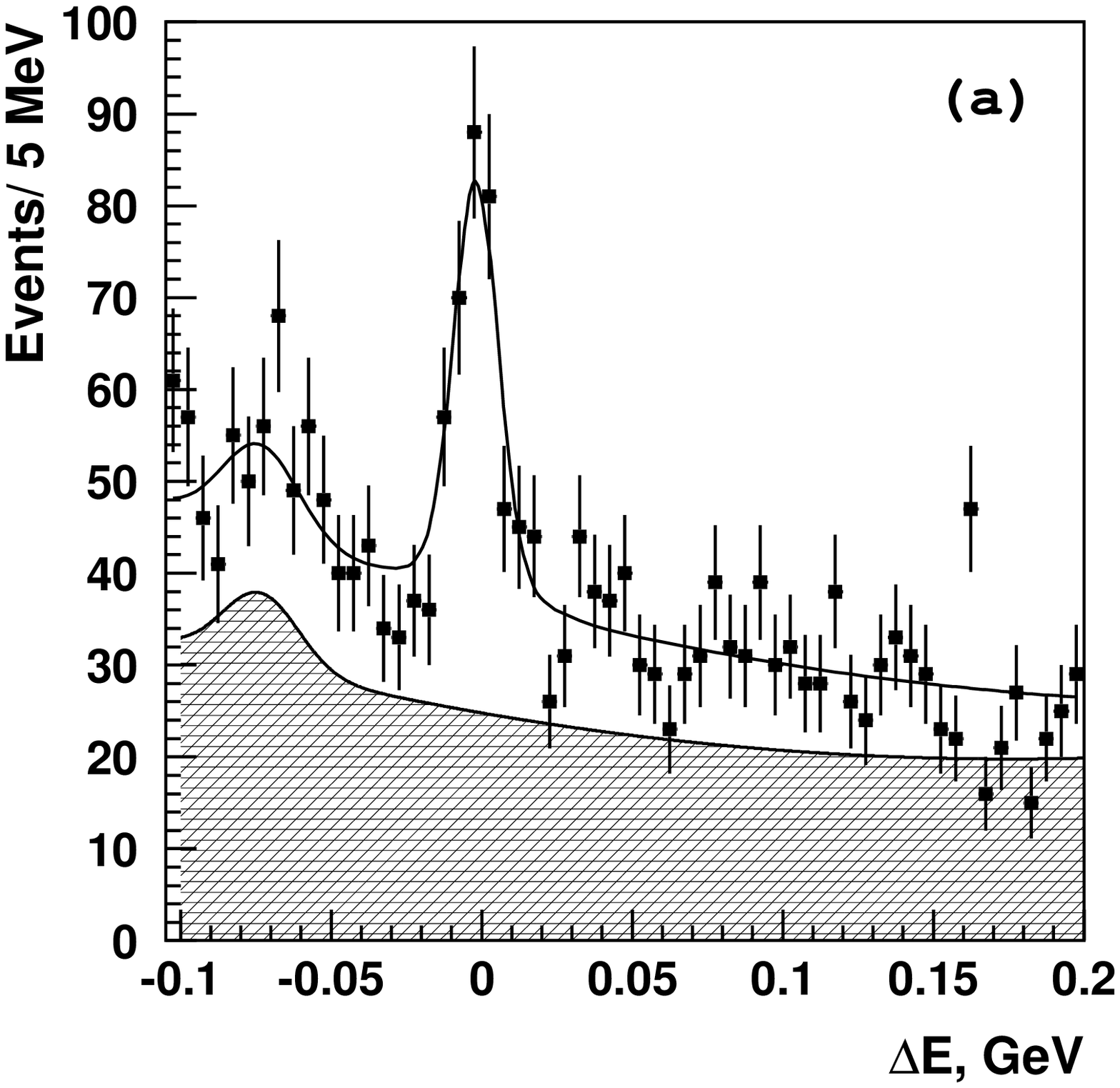}\hfill
  \includegraphics[width=0.39\textwidth] {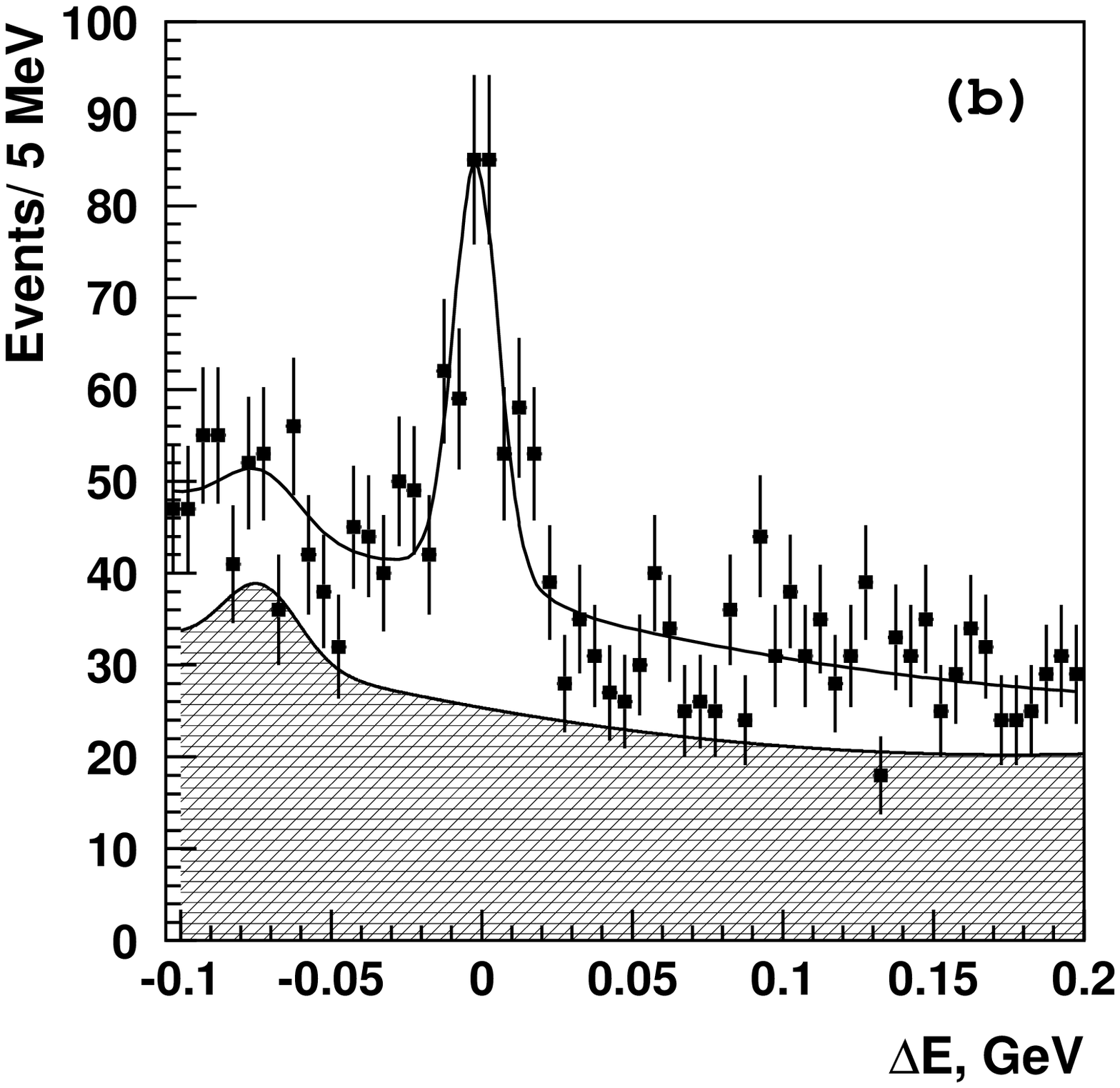}
  \caption{The $\de$ distribution for (a) $\bdndp$ and (b) $\bdndm$.
    Points with errors
    represent the experimental data, open curves show projections from the 
    2D fits and cross-hatched curves show the $\bb$ component only.}
  \label{dndc_de}
\end{figure*}

To calculate the charge asymmetry in the $\bdndp$ decay channel, we repeat 
the fits  separately for the $\bdndp$ and $\bdndm$ samples.
The $\de$ distributions for $\bdndp$ and $B^- \to D^- D^0$ candidates are 
presented in Fig.~\ref{dndc_de}.
The fit results are given in Table~\ref{asymfit}.
Using the results of the 2D fits, we obtain the charge asymmetry:
$$A_{CP}=\frac{N(\dndm)-N(\dndp)}{N(\dndm)+N(\dndp)}=\acpnbdp\ .$$

We calculate the $\bdndp$ branching fraction separately for each $D$ 
decay channel; the results are consistent with each other.
As an additional check, we apply a similar procedure to a decay chain 
with a similar final state: $B^+\to D_s^+\bar{D}^0$.
We measure the branching fraction 
$\br(B^+\to D_s^+\bar{D}^0)=(9.5\pm 0.2)\times 10^{-3}$,
where the error is statistical only. This is
consistent with the world average value 
$(10.0\pm 1.7)\times 10^{-3}$~\cite{PDG}.
The charge asymmetry in this final state is consistent with zero: 
$(-0.5\pm 1.5)\%$.
We also measure the charge asymmetry for the $D^+\bar{D}^0$ background 
events and find a value consistent with zero: $(-1.4\pm 1.3)\%$.

\begin{table}
\caption{Sources of systematic uncertainty.}
\medskip
\label{systematic}
\begin{tabular}{lccc}\hline\hline
  Source&$\br(D^0\bar{D}^0)$&$\br(D^+\bar{D}^0)$&$A_{CP}(D^+\bar{D}^0)$\\\hline
  $D$ branching fraction & $7\%$   & $5\%$   & $0$ \\
  Tracking           & $6\%$   & $6\%$   & $0.013$ \\
  PID                & $2\%$   & $4\%$   & $0.004$ \\
  $\pi^0$ reconstruction & $2\%$   & $1.3\%$ & $0$\\
  $N(B\bar{B})$      & $1.4\%$ & $1.4\%$ & $0$\\
  MC statistics      & $1\%$   & $1\%$   & $0$\\
  Signal fit         & $4\%$   & $4\%$   & $0.015$\\\hline
Total            & $10.5\%$ & $9.8\%$ & $0.02$\\\hline\hline
\end{tabular}
\end{table}

Table~\ref{systematic} shows the sources of the systematic uncertainty.
The errors due to knowledge of $D$ branching fractions are taken 
from Ref.~\cite{PDG}.
The uncertainty in the tracking efficiency is estimated using 
partially reconstructed $D^{*+}\to D^0[\ks\pi^+\pi^-]\pi^+$ decays. 
The uncertainty in the PID efficiency is determined from 
$D^{*+}\to D^0[K^-\pi^+]\pi^+$ decays.
The error in signal yield determination is estimated by varying the 
signal and background shapes and fit range.
We assume equal production rates for $B^+B^-$ and $B^0\bar B^0$ pairs 
and do not include the uncertainty related to this assumption in the 
total systematic error. 

The asymmetry measurement contains the following systematic errors: 
tracking efficiency difference for $\pi^\pm$ (0.013), 
particle identification efficiency difference for $\pi^\pm$ and $K^\pm$
(0.004) and signal yield determination (0.015). 
The total systematic uncertainty is 0.02.

In summary, we report improved measurements of the
$\bdndp$ branching fraction $\br(\bdndp)=\brdndp$.
The charge asymmetry for this decay is measured to be consistent with zero 
$A_{CP}(\bdndp)=\acpnbdp$.
We also set an upper limit for the $\bdndb$ decay branching fraction
of $\br(\bdndb)<\brdndb$ at 90\% CL.
These results are consistent with our previous
results~\cite{belle_dndp} and supersede them.
Our results are also consistent with BaBar measurements~\cite{babar_dndp}.

We thank the KEKB group for the excellent operation of the
accelerator, the KEK cryogenics group for the efficient
operation of the solenoid, and the KEK computer group and
the National Institute of Informatics for valuable computing
and Super-SINET network support. We acknowledge support from
the Ministry of Education, Culture, Sports, Science, and
Technology of Japan and the Japan Society for the Promotion
of Science; the Australian Research Council and the
Australian Department of Education, Science and Training;
the National Science Foundation of China and the Knowledge
Innovation Program of the Chinese Academy of Sciences under
contract No.~10575109 and IHEP-U-503; the Department of
Science and Technology of India; 
the BK21 program of the Ministry of Education of Korea, 
the CHEP SRC program and Basic Research program 
(grant No.~R01-2005-000-10089-0) of the Korea Science and
Engineering Foundation, and the Pure Basic Research Group 
program of the Korea Research Foundation; 
the Polish State Committee for Scientific Research; 
the Ministry of Education and Science of the Russian
Federation and the Russian Federal Agency for Atomic Energy;
the Slovenian Research Agency;  the Swiss
National Science Foundation; the National Science Council
and the Ministry of Education of Taiwan; and the U.S.
Department of Energy.

\end{document}